\documentclass{elsart}
\usepackage{epsfig}
    \newcommand{\ba}{\begin{eqnarray}}
    \newcommand{\ea}{\end{eqnarray}}
    \newcommand{\be}{\begin{equation}}
    \newcommand{\ee}{\end{equation}}

    \newcommand{\AmS}{{\protect\the\textfont2%
  A\kern-.1667em\lower.5ex\hbox{M}\kern-.125emS}}

\begin{document}
\runauthor{PKU and ITP and IHEP}
\begin{frontmatter}

\title{$I=2$ Pion Scattering Length from a Coarse Anisotropic Lattice %
Calculation %
 }

\author[PKU]{Chuan Liu}
\author[PKU]{Junhua Zhang}
\author[IHEP]{Ying Chen}
\author[ITP]{and J.~P.~Ma}
\address[PKU]{Department of Physics\\
          Peking University\\
                  Beijing, 100871, P.~R.~China}
\address[ITP]{Institute of Theoretical Physics\\
                Academia Sinica\\
                Beijing, 100080, P.~R.~China}
\address[IHEP]{Institute of High Energy Physics\\
                Academia Sinica\\
                P.~O.~Box 918\\
                Beijing, 100039, P.~R.~China}
                \thanks{This work is supported by the National
Natural Science Foundation of China (NSFC) and Pandeng fund.}

\begin{abstract}
Using the tadpole improved clover Wilson quark action on coarse
anisotropic lattices, the $\pi\pi$ scattering length in the $I=2$
channel is calculated within quenched approximation. We show that
such a calculation is feasible using small lattices on small
computers provided that the finite volume and finite lattice
spacing errors are under control. Our results are extrapolated
towards the chiral, infinite volume and continuum limit.
Comparisons of our results with previous lattice results from
JLQCD collaboration, the new results from E865 experiment, and the
results from Chiral Perturbation Theory are made. Good agreements
are found.
\end{abstract}
\end{frontmatter}


\section{Introduction}

 It has become clear that anisotropic, coarse lattices and improved
 lattice actions are ideal candidates for lattice
 QCD calculations on small computers %
 \cite{lepage93:tadpole,lepage95:pc,colin97,colin99}. They are
 particularly advantageous for heavy objects like the glueballs,
 one meson states with nonzero spatial momenta and
 multi-meson states with or without spatial momenta. The gauge action
 employed is the tadpole improved gluonic action on anisotropic
 lattices \cite{lepage95:pc}.
 Using this action, glueball and hadron spectrum has been studied
 within quenched approximation \cite{colin97,colin99,%
 chuan01:gluea,chuan01:glueb,chuan01:canton1,chuan01:canton2,chuan01:india}.

 In this letter, we report our results on
 the pion-pion scattering lengths within quenched
 approximation using relatively small lattices.
 Lattice calculations of pion scattering lengths
  have been performed by various authors
 using symmetric lattices without the
 improvement~\cite{gupta93:scat,fukugita95:scat,jlqcd99:scat}.
 With the symmetric lattices and Wilson
 action, large lattices (typically $24^364$ or larger)
 have to be used which require
 substantial amount of computer resources.
 It gets even more challenging if the
 chiral, infinite volume and continuum limit extrapolation is
 to be made~\cite{jlqcd99:scat}.
 In this letter, we show that such a calculation
 is feasible using relatively small
 lattices ($8^340$ or so) on small computers with the
 tadpole improved anisotropic lattice actions.
 Our final extrapolated result on the pion-pion scattering
 length in the $I=2$ channel is compared with previous lattice
 result~\cite{jlqcd99:scat}, the new
 experimental result from E865 collaboration and
 Chiral Perturbation Theory.

 The fermion action used in this calculation is the
 tadpole improved clover Wilson action on anisotropic
 lattices \cite{klassen99:aniso_wilson,chuan01:tune}.
 Among the parameters which appear in the fermion matrix,
 the so-called bare velocity of light $\nu$ is tuned non-perturbatively
 using the single pion dispersion relations \cite{chuan01:tune}.
 The tadpole improved tree-level
 values are used for other parameters.
 In the fermion matrix, the bare quark mass
 dependence is singled out so that one could utilize the
 shifted structure of the matrix to solve for
 quark propagators at various values of valance quark mass
 at the cost of solving only the lightest one,
 using the so-called  Multi-mass Minimal
 Residual ($M^3R$) algorithm %
 \cite{frommer95:multimass,glaessner96:multimass,beat96:multimass}.

 \section{Formulation to extract the scattering lengths}
 \label{sec:theory}

 In order to calculate hadron scattering lengths on the lattice,
 or the scattering phase shifts in general, one uses L\"uscher's
 formula which relates the exact energy level of two hadron states
 in a finite box to the scattering phase shift in the continuum.
 In the case of pion-pion scattering, this formula relates the
 {\em exact} two pion energy $E^{(I)}_{\pi\pi}$ in a finite
 box of size $L$ and isospin $I$
 channel to the corresponding scattering length
 $a^{(I)}_0$ in the continuum~\cite{luscher86:finiteb}:
 \be
 \label{eq:luescher}
 E^{(I)}_{\pi\pi}-2m_\pi=-\frac{4\pi a^{(I)}_0}{m_\pi L^3}
 \left[1+c_1\frac{a^{(I)}_0}{L}+c_2(\frac{a^{(I)}_0}{L})^2
 \right]+O(L^{-6}) \;\;,
 \ee
 where $c_1=-2.837297$, $c_2=6.375183$ are numerical constants.
 The formula suffers dramatic changes in the quenched
 approximation\footnote{%
 The authors would like to thank Professor
 C.~Bernard for bringing our attention to this issue.}%
 as discussed in \cite{bernard96:quenched_scat}.
 In the $I=0$ channel, one-loop quenched chiral perturbation
 theory gives anomalous contributions to the two-pion energy
 that are of order $L^0=1$ and of order $L^{-2}$ due to
 $\eta'$ loops. In the $I=2$ channel, these enhanced
 contributions are absent.

 To measure the pion mass $m_\pi$ and
 to extract the exact energy level $E^{(I)}_{\pi\pi}$
 of two pions with zero relative momentum, appropriate correlation
 functions are constructed from the corresponding operators
 in the $I=2$ channel.
 We have used the operators proposed in
 Ref.~\cite{fukugita95:scat}.
 Numerically, it is more advantageous to construct
 the ratio of the correlation functions defined above.
 It is argued \cite{sharpe92:scat,bernard96:quenched_scat}
 that one should use the linear fitting function:
 \be
 \label{eq:linear_fit}
 {\mathcal R}^{I=2}(t) \equiv C^{I=2}_{\pi\pi}(t) / (C_\pi(t)C_\pi(t))
  \stackrel{T >> t >>1}{\sim} 1-\delta E^{(2)}_{\pi\pi}t
 \;\;,
 \ee
 where $C^{I=2}_{\pi\pi}(t)$ and $C_\pi(t)$ are the two and one
 pion correlation functions and
 $\delta E^{(2)}_{\pi\pi}=E^{(2)}_{\pi\pi}-2m_\pi$
 is the energy shift.
 Two pion correlation function, or equivalently, the ratio
 ${\mathcal R}(t)$ constructed above can be
 transformed into products of quark propagators using
 Wick's theorem \cite{fukugita95:scat}.
 The $I=2$ two pion correlation function is given by
 two contributions which are termed Direct and
 Cross contributions \cite{gupta93:scat,fukugita95:scat}.
 The two pion correlation function in the
 $I=0$ channel is, however, more complicated which
 involves vacuum diagrams that require to compute
 the quark propagators for wall sources placed at
 {\em every} time-slice, a procedure which is more
 time-consuming than the $I=2$ channel.

 \section{Simulation details}
 \label{sec:simulation}

 Simulations are performed on several PCs and workstations.
 Configurations are generated using the pure
 gauge action with fixed anisotropy $\xi=5$
 for $4^340$, $6^340$ and $8^340$ lattices at the gauge
 coupling $\beta=1.7$, $2.2$, $2.4$ and $2.6$. The spatial lattice
 spacing $a_s$ is roughly between $0.19$fm and $0.39$fm while
 the physical size of the lattice ranges from
 $0.8$fm to $3.2$fm. For each set of parameters, several hundred
 decorrelated gauge field configurations are used to measure
 the fermionic quantities. Statistical errors are all analyzed
 using the usual jack-knife method.

 Quark propagators are measured using the Multi-mass Minimal
 Residue algorithm for $5$ different values of bare quark
 mass using  wall sources to enhance
 the signal \cite{gupta93:scat,fukugita95:scat,jlqcd99:scat}.
 Periodic boundary condition is applied to all three
 spatial directions while in the temporal direction, Dirichlet
 boundary condition is utilized.

 The single pseudo-scalar and vector meson
 correlation functions at zero spatial momentum and three lowest lattice
 momenta, namely $(100)$, $(110$ and $(111)$ are constructed from
 the corresponding quark propagators.
 Using the anisotropic lattices, we are
 able to obtain decent effective mass plateaus for these energy
 levels. The parameter $\nu$, also known as the bare velocity of light,
 that enters the fermion matrix is determined non-perturbatively
 using the single pion dispersion relations as described in
 Ref.~\cite{chuan01:tune}.

 Two pion correlation functions and the ratio ${\mathcal R}(t)$ are
 constructed for all $\kappa$, $\beta$ and $L$ values
 as a function of the temporal separation $t$.
 Then, the linear fit~(\ref{eq:linear_fit}) is performed.
 Again, due to the usage of the anisotropic lattices,
 we obtained reasonable signal for the energy shift with
 a typical error around ten percent.
 The energy shifts $\delta E^{(2)}_{\pi\pi}$ are
 substituted into L\"uscher's formula to solve
 for the scattering length $a^{(2)}_0$ for a given set of
 parameters. From these results, we could perform
 an extrapolation towards the chiral, infinite volume
 and zero lattice spacing limit.

 In the chiral limit, the $\pi\pi$ scattering length in
 the $I=2$ channel is given by the current algebra
 result due to Weinberg~\cite{weinberg:scat} in
 {\em full} QCD:
 \be
 \label{eq:weinberg}
 a^{(2)}_0 =-{1 \over 16 \pi} {m_\pi \over f^2_\pi} \;\;,
 \ee
 where $f_\pi\sim 93$MeV is the pion decay constant.
 Chiral Perturbation Theory results to one-loop and two-loop
 order have been calculated~\cite{leutwyler83:chiral,bijnens:scat}.
 However, the one-loop and two-loop {\em numerical}
 results on the pion-pion
 scattering length in the $I=2$ channel do not differ from
 the current algebra value substantially.
 Complication arises in the quenched approximation. In principle,
 the quenched scattering lengths becomes divergent in the
 chiral limit \cite{bernard96:quenched_scat}. However, these
 divergent terms only become numerically important when the
 pion mass is close to zero. For the parameters used in our
 simulation, these terms seem to be numerically small in the
 $I=2$ channel and we could not observe any sign of
 divergence from our data.

 In early lattice
 calculations \cite{gupta93:scat,fukugita95:scat,jlqcd99:scat},
 both the mass and the decay constant of the pion
 were calculated on the lattice. Then, the lattice
 results of $m_\pi$ and $f_\pi$ were substituted
 into the current algebra result~(\ref{eq:weinberg})
 to obtain a prediction of the scattering length.
 This is to be compared with the scattering length
 obtained from the energy shifts on the
 lattice and L\"uscher's formula. In these studies,
 some discrepancies between the lattice results and
 the chiral results were observed.
 The disadvantage
 of the this procedure is the following: first, it is
 difficult to make a direct chiral extrapolation;
 second, the lattice results of the
 decay constant are usually much less accurate,
 both statistically and systematically, than the
 mass values. In fact, most of the discrepancies are
 due to inaccuracy of the decay constants, as
 realized in ~\cite{jlqcd99:scat}.
 In Ref.~\cite{jlqcd99:scat}, the authors proposed to
 use the quantity $a^{(2)}_0/m_\pi$, which is much better
 behaved in the chiral limit. They found that the results
 obtained using this quantity is in much better agreement
 with both current algebra and the experiment.
 In this letter, we use a similar but dimensionless
  quantity $F=a^{(2)}_0m^2_\rho/m_\pi$, which in
 the chiral limit reads:
 \be
 \label{eq:chiral}
 F\equiv {a^{(2)}_0m^2_\rho \over m_\pi}
 =-{1 \over 16 \pi} {m^2_\rho \over f^2_\pi}
 \sim  -1.3638\;\;,
 \ee
 where the final numerical value is obtained
 by substituting in the experimental values for
 $m_\rho$ and $f_\pi$. On the lattice, the scattering
 length $a^{(2)}_0$  is extracted from L\"uscher's formula.
 More importantly,
 the mass of the pion and the rho can be obtained with
 {\em good} accuracy on the lattice. So, the factor $F$
 can be calculated on the lattice with good
 precision {\em without} the
 lattice calculation of $f_\pi$. The error
 of the factor $F$ obtained on the lattice will mainly
 come from the error of the scattering length $a^{(2)}_0$, or
 equivalently, the energy shift $\delta E^{(2)}_{\pi\pi}$.
 Since we have calculated the factor $F$ for $5$ different
 values of valance quark mass, we could perform a chiral
 extrapolation and extract the result of $F$ in the chiral limit.
 Comparisons with Weinberg's result~(\ref{eq:chiral}) and
 the experiment will offer us a cross check among different
 methods.

 \begin{figure}[thb]
 \begin{center}
 \includegraphics[height=9.0cm,angle=0]{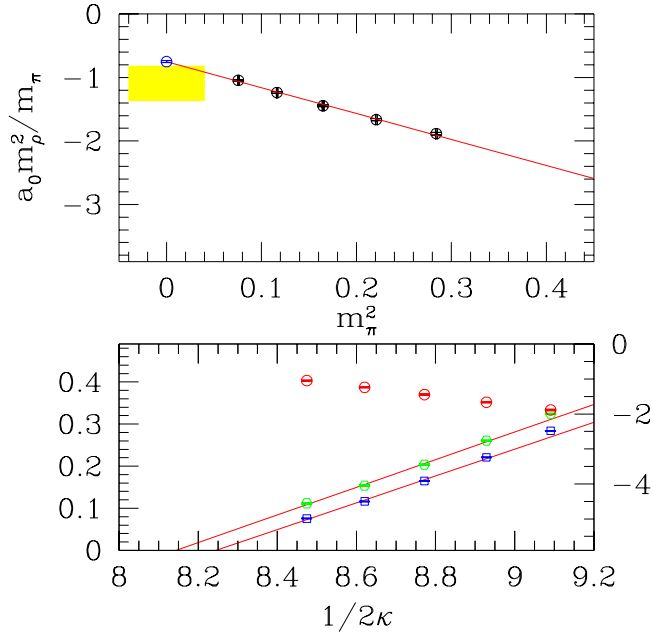}
 \end{center}
 \caption{Chiral extrapolation for the quantity
 $F$ for our simulation results
 at $\beta=2.2$ on $4^340$ lattices. In the lower
 half of the plot, the pseudo-scalar and vector
 meson mass squared are plotted as open squares and
 open hexagons, respectively, as functions of
 $1/(2\kappa)$. The straight lines represent
 the corresponding linear fit for them.
 Also shown in the lower half as open circles are the results
 for the factor $F$. In the upper half
 of the plot, the same quantity $F$ is shown
 as a function of $m^2_\pi$. The straight line shows the linear
 extrapolation towards the chiral limit $m^2_\pi=0$, where
 the extrapolated result is also shown. As a comparison, the
 corresponding experimental result \cite{experiment_new1:scat}
 for this quantity is shown as a shaded band.
 \label{fig:chiral_extrapolation}}
 \end{figure}
 As an illustration in the
 upper half of Fig.~\ref{fig:chiral_extrapolation},
 we show the chiral extrapolation of the quantity
 $F$ as a function of the pseudo-scalar
 mass squared ($m^2_\pi$) for the simulation
 on  $4^340$ lattices at $\beta=2.2$.
 The fitting quality for the chiral extrapolation of our
 data at other simulation parameters
 are quite similar. The pseudo-scalar
 and vector meson mass squared $m^2_\pi$ and $m^2_\rho$
 are also shown in the lower half of the figure as
 a function of $1/(2\kappa)$, which linearly depends
 on the valance quark mass.
 It is seen that meson mass squared depends on
 the valance quark mass linearly. The data points
 for the factor $F$ are also shown in the lower half
 of the plot. The fitting quality for the pion, rho
 and the factor $F$ is reasonable. Admittedly, it is a bit astonishing
 to observe such a linear behavior since our data
 were obtained at relatively heavy pion mass values.
 Typical $m_\pi/m_\rho$ values are above $0.7$.
 We do not have a theoretical explanation for
 this behavior at the moment. Using our data, we have
 tried to make the chiral extrapolation by adding a term
 proportional to $m^4_\pi$ to our fitting function. However, this
 does not improve the quality of the fit and the fitted coefficient
 of the $m^4_\pi$ term turns out to be consistent with zero with a
 large error. Therefore, in the following we will only quote
 our results using a simple linear extrapolation.

 \begin{figure}[thb]
 \begin{center}
 \includegraphics[height=9.0cm,angle=0]{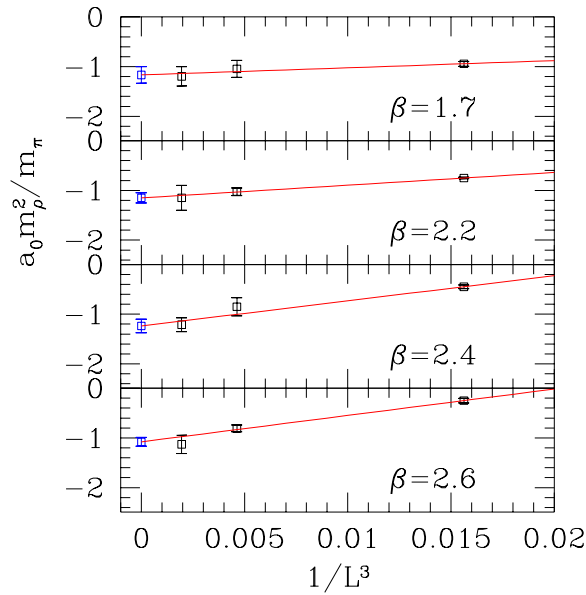}
 \end{center}
 \caption{Infinite volume extrapolation in Scheme I for the quantity
 $F=a^{(2)}_0m^2_\rho/m_\pi$ for our simulation results
 at $\beta=2.6$, $2.4$, $2.2$ and $1.7$. The straight lines represent
 the corresponding linear extrapolation in $1/L^3$.
 The extrapolated result is also shown, together with its error.}
 \label{fig:extrapL}
 \end{figure}
 After the chiral extrapolation, we now turn to study
 the finite volume effects of the simulation. According
 to formula~(\ref{eq:luescher}), the quantity
 $F$ obtained on finite lattices differ from
 its infinite volume value by corrections of the
 form $1/L^3$. However, it was argued in
 Ref.\cite{sharpe92:scat,fukugita95:scat,bernard96:quenched_scat}
 that in a quenched calculation, the form of L\"uscher's
 formula is invalidated by finite volume corrections
 of the form $1/L^5$ instead of $1/L^6$. This would mean that the
 factor $F$ receives finite volume
 correction of the form $1/L^2$. In our simulation, however,
 we were unable to judge from our data which extrapolation
 is more convincing. We therefore performed our infinite volume
 extrapolation in both ways, calling them scheme I (extrapolating
 according to $1/L^3$) and scheme II (extrapolating according to
 $1/L^2$), respectively.
 Extrapolation in these two different
 schemes yields compatible results within
 statistical errors. The fitting quality of Scheme II
 is somewhat, but not overwhelmingly, better than that of
 Scheme I.
 In Fig.~\ref{fig:extrapL},
 we show the infinite volume extrapolation according to
 Scheme I for the simulation points
 at $\beta=2.6$, $2.4$, $2.2$ and $1.7$. The extrapolated
 results are shown with blue open squares at
 $L=\infty$, together with the corresponding errors. The straight lines
 represent the linear extrapolation in $1/L^3$.
 It is seen that, on physically small lattices, e.g. those with
 $\beta=2.6$ and $\beta=2.4$, the finite volume correction
 is much more significant than larger lattices.
 This is also reflected by
 the slopes of the linear fits. The infinite volume
 extrapolation in Scheme II is similar.

 \begin{figure}[thb]
 \begin{center}
 \includegraphics[height=9.0cm,angle=0]{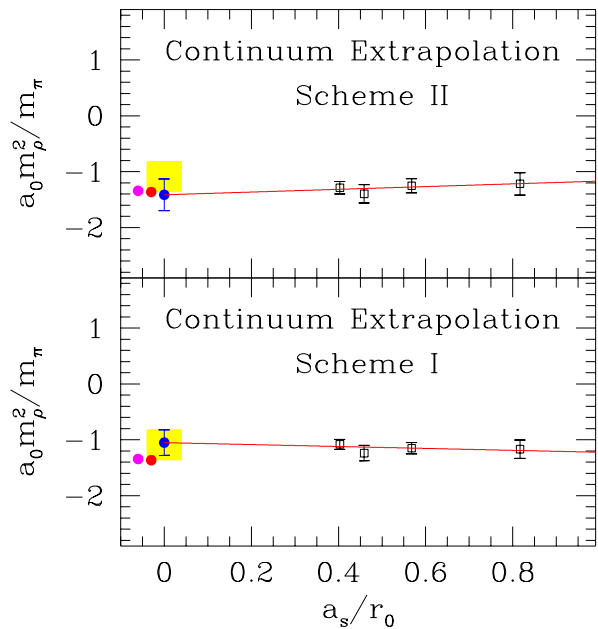}
 \end{center}
 \caption{Continuum extrapolation for the quantity
 $a^{(2)}_0m^2_\rho/m_\pi$ obtained from our simulation results
 at $\beta=2.6$, $2.4$, $2.2$ and $1.7$. The results for both
 Scheme I and II are shown. The straight lines represent
 the linear extrapolation in $a_s/r_0$.
 The extrapolated results are also shown, together with the
 experimental result from Ref.~\cite{experiment_new1:scat} indicated
 by the shaded band. For comparison, Weinberg's
 result~(\ref{eq:chiral}) and the results from Chiral perturbation
 theory are also shown as the red and magenta points respectively.
 \label{fig:continuum_extrapolation}}
 \end{figure}
 Finally, we can make an extrapolation towards the
 continuum limit by eliminating the finite lattice
 spacing errors. Since we have used the tadpole
 improved clover Wilson action, all physical quantities
 differ from their continuum counterparts by
 terms that are proportional to $a_s$. The physical
 value of $a_s$ for each value of $\beta$ can be
 found from Ref. \cite{colin99,chuan01:india}.
 This extrapolation is shown in
 Fig.~\ref{fig:continuum_extrapolation} where
 the results from the chiral and infinite volume
 extrapolation discussed above are indicated as
 data points in the plot for all $4$ values of
 $\beta$ that have been simulated. In the lower/upper half
 of the plot, results in Scheme I/II are shown.
 The straight lines show
 the extrapolation towards the $a_s=0$ limit and the
 extrapolated results are also shown together with
 the experimental result from Ref.\cite{experiment_new1:scat}
 which is shown as the shaded band. For comparison with
 chiral perturbation theory, Weinberg's
 result~(\ref{eq:chiral}) and the results from Chiral perturbation
 theory are also shown as the red and magenta points respectively.
 It is seen that our lattice calculation gives a compatible result
 for the quantity $F$ when compared
 with the experiment. The statistical error of our final
 result is about the same as that of Ref.~\cite{jlqcd99:scat}.
 We think the result is promising since we have shown that
 such a calculation can be obtained on relatively small lattices
 and limited computer resources.
 Another encouraging sign
 is that all data points, even the one at
 a lattice spacing of $0.4$fm, show little dependence on $a_s$
 indicating that the $O(a_s)$ lattice
 effects are small, presumably due to the tadpole improvement of
 the action. The slope of our linear extrapolation is small.
 This is very helpful in cutting down the error of
 the final extrapolated result. In Ref.~\cite{jlqcd99:scat},
 continuum extrapolation was also performed for Wilson fermions without
 the improvement. There, a larger slope of the linear extrapolation is
 seen, see Fig.~4 of Ref~\cite{jlqcd99:scat}.
 This magnifies the error of the final
 extrapolated result when compared with
 the errors of the results at finite lattice spacings.

 To summarize, we obtain from the linear continuum
 extrapolation the following result for the quantity $F$:
 \ba
 {a^{(2)}_0m^2_\rho \over m_\pi}&=&-1.05(23)
 \;\;\mbox{for Scheme I}\;\;,
 \nonumber \\
 {a^{(2)}_0m^2_\rho \over m_\pi}&=&-1.41(28)
 \;\;\mbox{for Scheme II}\;\;.
 \ea
 If we substitute in the mass of the mesons from
 the experiment, we obtain the quantity $a^{(2)}_0m_\pi$:
 \ba
 a^{(2)}_0m_\pi&=&-0.0342(75) \;\;\mbox{for Scheme I}\;\;,
 \nonumber \\
 a^{(2)}_0m_\pi&=&-0.0459(91) \;\;\mbox{for Scheme II}\;\;.
 \ea
 This result is compatible with previous lattice result
 using Wilson fermions on large lattices~\cite{jlqcd99:scat}.
 Current algebra prediction~(\ref{eq:weinberg})
 yields a value of $a^{(2)}_0m_\pi=-0.046$.
 This quantity has been calculated in Chiral Perturbation
 Theory to one-loop order with the result:
 $a^{(2)}_0m_\pi=-0.042$ \cite{leutwyler83:chiral} and
 recently to two-loop order \cite{bijnens:scat,leutwyler01:scat}.
 The final result from Chiral Perturbation Theory
 gives: $a^{(2)}_0m_\pi=-0.0444(10)$, where the
 error comes from theoretical uncertainties.
 On the experimental side, a new
 result from E865 collaboration \cite{experiment_new1:scat}
 claims $a^{(2)}_0m_\pi=-0.036(9)$. It is encouraging to find out
 that our lattice results in {\em both} schemes are compatible with
 the experiment. Our result in Scheme II also agrees with the
 Chiral Perturbation Theory results very well while our result
 in Scheme I is barely within one standard deviation of the chiral
 results.

 \section{Conclusions}
 \label{sec:conclude}

 In this letter, we have calculated pion-pion scattering
 lengths in isospin $I=2$ channel using quenched
 lattice QCD. It is shown that such a calculation
 is feasible using coarse, anisotropic, small lattices
 with limited computer resources like several
 personal computers and workstations.
 The calculation is done using the
 tadpole improved clover Wilson action on anisotropic
 lattices. The anisotropy helps to enhance the temporal
 resolution of correlation functions while the improvement
 helps to cut down the finite lattice spacing errors on
 coarse lattices. Simulations are performed on lattices
 with various sizes, ranging from $0.8$fm to about
 $3$fm and with different value of lattice spacing.
 The infinite volume extrapolation is explored in
 two different schemes which yields compatible final
 results. The lattice result for the scattering length is
 extrapolated towards the chiral
 and continuum limit where a result consistent with
 the experiment is found. Comparisons with previous lattice
 results and the Chiral
 Perturbation Theory are also made with encouraging results.
 Our data and results also suggest that,
 using improved gluonic and fermionic actions, the lattice
 spacing errors for the scattering length is under control
 even on coarse lattices of $a_s\sim 0.4$fm.

 Finally, our method for calculating the pion-pion
 scattering length discussed here can be easily
 generalized to calculate the scattering lengths
 of other hadrons, or in other channels, e.g. $I=0$ channel
 where extra care has to be taken due to enhanced terms
 coming from quenched chiral loops.
 The method can also be applied to calculate the scattering
 phase shift at non-zero spatial lattice momenta, where
 presumably larger lattices are needed.


\end{document}